\documentclass[10pt,conference]{IEEEtran}
\usepackage{spconf,amsmath,graphicx,cite}
\usepackage{amsfonts,amssymb}
\usepackage{color}
\usepackage[french]{babel}

\usepackage[absolute]{textpos}

\usepackage{algorithm}
\usepackage{algpseudocode}

\usepackage{etoolbox}

\algrenewcommand\alglinenumber[1]{{\sffamily\footnotesize#1}}

\makeatletter
\expandafter\patchcmd\csname\string\algorithmic\endcsname{\itemsep\z@}{\itemsep=1ex plus1pt}{}{}
\makeatother

\title{A B\MakeLowercase{ootstrap} M\MakeLowercase{ethod for} S\MakeLowercase{inusoid} D\MakeLowercase{etection in} C\MakeLowercase{olored} N\MakeLowercase{oise and} U\MakeLowercase{neven} S\MakeLowercase{ampling}. \\ A\MakeLowercase{pplication to} E\MakeLowercase{xoplanet} D\MakeLowercase{etection}}

\name{Sophia Sulis, David Mary, Lionel Bigot}
\address{Universit\'e C\^ote d'Azur, Observatoire de la C\^ote d'Azur, Laboratoire Lagrange, \\     
        UMR CNRS 7293, CS 34229, 06304 Nice, France}

\begin{document}

\begin{textblock}{15.5}(0.2,0.2)
\noindent  \centering First published in the Proceedings of the 25th European Signal Processing Conference (EUSIPCO-2017) in 2017, published by EURASIP.\\
\vspace{0.3cm}
\hrule
\end{textblock}

\maketitle

\begin{abstract}
This study is motivated by the problem  of  evaluating reliable false alarm (FA) rates for sinusoid detection tests applied to unevenly sampled time series involving 
 colored noise, when a (small) training data set of this noise is available.
While analytical expressions for the FA rate are out of reach in this situation, we show that it is possible to combine 
specific periodogram standardization and  bootstrap techniques to consistently estimate the FA rate. We also show that the procedure can be improved by using generalized extreme-value distributions. The paper presents several  numerical results including a case study in exoplanet detection from radial velocity data.
\vspace{-0.05cm}\end{abstract}



\section{Introduction}
\label{sec:intro}

 This study is motivated by the problem of assessing reliable false alarm (FA) rates for detection tests when the noise is colored, with partially unknown statistics and when the sampling is not regular.  Sinusoid (or more generally periodicity) detection tests are often based on the (Schuster's) periodogram \cite{Schuster_1898}, defined as
\begin{equation} 
\label{eq1}
	P(\nu):=   \frac{1}{N}  \Big| \sum_{j=1}^{N}  X(t_j)\mathrm{e}^{-i2\pi\nu t_j} \Big|  ^2,
	\vspace*{-0.1cm}
\end{equation}
with $\{X(t_j)\}_{j=1,\hdots,N}$  the data samples and $\nu$ the  frequency. \\
In the case of uneven time sampling, which is of concern in the present study, the periodogram ordinates in \eqref{eq1} are not independent, even when the noise is white Gaussian,  because the sampled exponentials are not orthogonal in general \cite{Scargle_1982}. This dependence makes the analysis of FA rates difficult.\\
In the literature of periodicity detection, numerous ``generalized'' periodograms (i.e., different from \eqref{eq1}) have been proposed (\textit{e.g.}, \cite{Scargle_1982,Zechmeister_2009}),
which correlate the time series with (possibly highly redundant) dictionaries of target signatures. As for Schuster's periodogram however, owing to at least one of the three following factors :\\ (i) uneven sampling, \\(ii) intrinsic correlation of the considered features,  \\(iii) partially unknown  noise correlations or dependencies,\\  the
joint  distribution of these periodogram ordinates is unknown (although the marginal distribution can be obtained in some cases), which prevents from deriving accurate false FA rates  (see \cite{Suveges_2012}).
Empirical methods exist (\textit{e.g.},\cite{Horne_1986}) but they often lead to inconsistencies \cite{Frescura_2008, Suveges_2012}.  To the best of our knowledge, the problem of accurately estimating the FA rate of periodograms-based detection tests in the case of uneven sampling and partially unknown colored noise remains open. Yet, the case of irregular sampling is frequently encountered in practice, for instance in astronomical observations \cite{Scargle_1982,Suveges_2012}. \\
The present study explores a bootstrap solution to this problem. As in \cite{Sulis_2017a}, we assume that  a training data set ${\cal{T}}_L$ of the noise is available. However, this set provides only a {\textit{small}} number of noise samples -  comparable typically to the number of data samples. In this regime, our knowledge of the statistics of the random process under ${\mathcal{H}}_0$ remains strongly impacted by estimation noise. The working hypothesis of an available training data set is rather general, but arises in particular  in the field of exoplanet detection in Radial Velocity (RV) data, where astrophysical models allow to generate time series of the noise - yet at a computational cost that makes long simulated time series out of reach. We show in the end of the paper that  accurate estimates of the FA rate can nevertheless be obtained in such conditions, at least for some tests.  To do so, our strategy will be to cancel the nuisances caused by the unknown noise statistics by considering standardized periodograms, for which the $\{t_j\}$ in \eqref{eq1} are unevenly spaced, and to capture the dependencies between the periodogram ordinates through bootstrap techniques.

In essence, the bootstrap is a computational tool for statistical inference using sample-driven Monte Carlo (MC)  simulations to produce empirical estimates of several statistical quantities as, here, the periodogram distribution  and estimates of the FA rates\cite{Zoubir_2004}.
These techniques, initially introduced for independent and identically distributed random variables (r.v.), have been intensively studied in the last decades  \cite{Zoubir_2004}.   
Bootstrap techniques exist in the case of even sampling and weakly dependent data (\textit{e.g.},
  \cite{Paparoditis_1999}) and different resampling procedures have been proposed (\textit{e.g.},\cite{Politis_1994}). 
For longer memory processes,  autoregressive  (AR)-aided  periodogram bootstrap  consists in estimating the parameters of an AR model in the time domain and in applying a non-parametric kernel based correction in the frequency domain to counteract the effects of estimation noise on the AR parameters (\textit{e.g.}, \cite{Kreiss_2003}).
For the uneven sampling case,  ``dependent wild bootstrap'' adapted to weakly dependent data unevenly sampled can be found in \cite{Shao_2010} and a bootstrap procedure using a Generalized Extreme Value (GEV) fit on the periodogram maxima is presented in \cite{Suveges_2012}. 
As underlined in \cite{Young_1994} for instance, the choice of the parameters of such procedures (\textit{e.g.}, parametric noise models, number of blocks, block lengths, data weights or repartition of the sampled blocks) can be delicate.

\noindent To conclude this short review of a large literature, controlling  the FA rate in sinusoid detection is problematic in the case of an uneven sampling and partially unknown colored noise, even when resorting to bootstrap techniques. 

In this study, we renounce to obtain analytical expressions of the $P_{FA}$. The contributions are the following: in Sec.\ref{sec1} and Sec.\ref{sec2}, we propose an original and automated bootstrap method based on a parametric-aided periodogram allied to periodogram standardization.
In Sec.\ref{sec3}, we propose to take advantage from GEV distributions to improve the method and, in Sec.\ref{sec4}, we present some of our numerical results, including an application to exoplanet detection.

\section{Standardized periodogram and statistical test}
\label{sec1}
Let us consider an observed time series of Power Spectrum Density (PSD) $S$, noted $X_{obs}|S$, involving a random process (with unknown PSD $S_E$), plus possibly a periodic component, sampled at uneven time instants $\{t_j\}_{j=1,\hdots,N_{obs}}$.
In some applications a training data set of the noise is available.
Specifically, let us assume that this training data set $ {\cal{T}}_L$ is composed with $L$  time series of the noise  $ \{X_\ell \}_{\ell = 1,\hdots,L}$, sampled at the same time instants $\{t_j\}$.
This set can be exploited to improve the control of the FA rate.
In this aim, the study \cite{Sulis_2017a} considered a standardized periodogram of the form:
\begin{equation}
\label{eq_ptilde0}
	 \widetilde{P}(\nu_k; X_{obs}|S_E, \{X_\ell|S_E\} ) := \frac{P(\nu_k;X_{obs}|S)}{\overline{ P}_L(\nu_k;\{X_\ell|S_E\})}
	 \vspace{-0.3cm}
\end{equation}
where 	
\begin{equation} 
\label{eq2}
	\overline{P}_L(\nu_k): =  \frac{1}{L} \sum_{\ell=1}^{L}  \frac{1}{N} \Big| \sum_{j=1}^{N} X_\ell(t_j)\mathrm{e}^{-i2\pi\nu_k t_j} \Big| ^2
	\vspace*{-0.1cm}
\end{equation}
is an estimate of the noise PSD. In this setting, \cite{Sulis_2017a} shows that $\widetilde{P}$ is independent of the unknown noise PSD and $F(2,2L)$-distributed. The stochastic estimation noise on the PSD $S_E$ is encapsulated by the  parameter $L$, with of course better detection performances for larger $L$.
Several CFAR detectors can be derived using  this standardized periodogram, whose FA rate can be analytically characterized in the case of even time sampling \cite{Sulis_2017a}.  In the present study, we focus on uneven sampling. We illustrate our study with $P$ as defined by \eqref{eq1} but other periodograms could be used, \textit{e.g.}, \cite{Scargle_1982,Zechmeister_2009}.
 Because of lack of space we consider below only one test, namely the test of the maximum applied to standardized ordinates $[ \widetilde{P}_{\nu_1} \hdots  \widetilde{P}_{\nu_K}]$. 

\noindent This test is defined as
\vspace{-0.2cm}
\begin{equation}
\label{eq_TM}
	M ({\widetilde{P}})\mathop{\gtrless}_{\mathcal{H}_0}^{\mathcal{H}_1} \gamma, \quad{\rm{with}}\quad
	M({\widetilde{P}}) := \displaystyle{\max_k} ~ \widetilde{P}(\nu_k),
\vspace{-0.3cm}
\end{equation}
and $\gamma$ is the detection threshold.
The associated $P_{FA}$ is defined as:
$$
P_{FA}(\gamma; M({\widetilde{P}})) := \textrm{Pr}\; (M({\widetilde{ P}})\! > \!\gamma | {\cal{H}}_0\!).
$$


\vspace*{-0.3cm}
\section{Proposed bootstrap procedure}
\label{sec2}
Our goal is to obtain a confidence interval for the $P_{FA}$ corresponding to any value of $\gamma$.
Of course, if we could generate as many realizations as desired of the r.v. $M(\widetilde{P})$ (resulting in, say $b$, realizations $\{m_i\}_{i=1,\hdots,b}$ of the r.v. $M$, collected in a vector ${{\bf{m}}}:=[m_1\hdots m_b]^\top$), we would be able to estimate the empirical distribution of the maximum and thus the $P_{FA}$ by:
\begin{equation}
	\widehat{P}_{FA}(\gamma; {\bf m}) := 1 - \widehat{\Phi}_{M}(\gamma; {\bf m}),
	\label{hatpfa}
\end{equation}
where $\widehat{\Phi}_{M}$ is the empirical cdf of $M$, and where the dependency of the estimates on the observed values $\bf{m}$ is explicit.\\ We can not do so however, because the training data set has finite size $L$. To counteract this fact, we may estimate the parameters of the noise from $\{X_\ell|S_E\}_{\ell=1,..,L}$ under
some model, with the aim of generating secondary noise data sets according to this estimated model. We opt for AR models, with PSD:
\vspace{-0.2cm}
\begin{equation}
	S_{E,AR}(\nu;{\boldsymbol{\theta}}_{AR})\!: =\! \frac{\sigma^2_{o}}{\Big|1\!+\!\displaystyle{\sum_{j=1}^{o}} c_j \mathrm{e}^{-2\pi i j \nu }\! \Big|^2},	 	
	\label{eq_SAR}
	\vspace{-0.3cm}
\end{equation}
where ${\boldsymbol{\theta}}_{AR}: = [o\;c_j\; \sigma_o^2]^\top$ is the AR parameter vector (order $o$, coefficients $\{c_j\}_{j=1,\cdots,{o}}$ and prediction error variance $\sigma_o^2$).
Various selection criteria exist to estimate the order ({\textit{e.g.}}, \cite{Akaike_1969, Rissanen_1984, Ding_2015}). We consider here the Bridge criterion ($BC$)\cite{Ding_2015}, known to mix the advantages of the Akaike and Bayesian information criteria:
 \vspace{-0.3cm}
 \begin{equation}
\!\! \widehat{o}=\arg\min_{{o}} {\rm BC}({o})~~{\rm with}~~{\rm BC}({o})\!:= \!\log{{\sigma}^2_{{o}}}\! +\! 2 \frac{o_{M}}{N}\displaystyle{\sum_{i=1}^{ {o}}} \frac{1}{i},
	\label{eq_o}
 \vspace{-0.2cm}
 \end{equation}
 with $o_{M}$ the largest candidate order and $\widehat{{\boldsymbol{\theta}}}_{AR} := [\widehat{o}\; \widehat{c}_j\; \widehat{\sigma}^2_{\widehat{o}}]^\top $ the estimated AR parameters, from which a PSD estimate $\widehat{S}_{E}$ can be obtained as in \eqref{eq_SAR}.
These parameters also allow to generate $(L+1) \times b$ correlated time series, noted $\{X_\ell | \widehat{S}_{E} \}$, allowing to generate simulated standardized periodograms $
\{ \widetilde{P}(\nu_k;X| \widehat{S}_{E}, \{X_{\ell} |\widehat{S}_{E}\})\}$, each requiring $1$ time series for numerator \eqref{eq1} and $L$ for denominator \eqref{eq2}. The resulting $b$ maxima can be used to estimate the ${P}_{FA}$ with \eqref{hatpfa}. 

However, this estimate is obtained by means of {\textit{one}} set of AR parameters estimated from ${\cal{T}}_L$. Obviously, we would have obtained a different estimate for a different training set ${\cal{T}}_L$. The question is therefore to know the distribution of this estimate, which we call ${\cal{D}}_{\widehat{P}_{FA}}$ below. The knowledge of this distribution would allow to provide the desired confidence interval. Obtaining ${\cal{D}}_{\widehat{P}_{FA}}$ is not straightforward, however, as only one genuine training set is available. This is where the bootstrap really comes in : we propose to use the estimated AR coefficients $\{\widehat{c}_j\}$ to generate a number $B$ of fake training data sets, from which we can obtain an estimate of this distribution, $\widehat{\cal{D}}_{\widehat{P}_{FA}}$. 
 This leads to the bootstrap procedure, called $\textsf{B}_0$, described in the Algorithm.
It begins by a first AR estimation on ${\cal{T}}_L$ (line 2).
 Secondary PSD estimates (noted $\widehat{\widehat{S}}_{E}^{(i)}, i=1,\hdots,B$) are generated
by reestimating the AR parameters on a $i$-th fake training data set noted $\{X_\ell |\widehat{S}_{E}\}$ (for simplicity, we omit the dependence in $i$ in this notation, lines 4-5). These parameters are used to generate $b$ standardized periodograms (line 7-11), the corresponding maxima (line 12) and $P_{FA}$ estimates (line 14). The distribution $\widehat{\cal{D}}_{\widehat{P}_{FA}}$ is estimated from the set of $B$ such estimates, $\{\widehat{P}_{FA}^{(i)}\}_{i=1,..,B}$ (line 16).
\noindent Of course, one advantage of this procedure is that it is independent from the data under test (in contrast to procedures where training data set are not available, and for which the noise statistics must be obtained from the data \cite{Priestley_1981}). 
An other interesting point is that the marginal distribution of $\widetilde{P}(\nu_k)$ can be estimated for all $k$, opening the door to tests other than \eqref{eq_TM} (see \cite{Sulis_2017a}). 
We note, however, that the choice of the parametric model has to be done very cautiously. As in \cite{Suveges_2012}, diagnostic plots can be used for model validation.
\vspace{-0.2cm}
\begin{algorithm}
 	\caption{Proposed bootstrap procedure ($\textsf{B}_0$) }
 	 \label{algo1}
 	\begin{algorithmic}[1]
 		\Procedure{$\textsf{B}_0$~}{${\cal{T}}_L$} 
		 	 \State $\widehat{S}_E(\nu_k;\widehat{\theta}_{AR}(\{X_\ell|S_E\}))$ \Comment{1$^{st}$ PSD estimation \eqref{eq_o}} 
 		 	 \For{$i = 1,\hdots, B$}	
			 	\State $ X_\ell |\widehat{S}_{E}$ \Comment{$L$ time series with PSD $\widehat{S}_{E}$}
				 \State $\widehat{\widehat{S}}_{E}^{(i)}(\nu_k;\widehat{\widehat{\theta}}_{AR}(\{X_\ell |\widehat{S}_{E}\}$) \Comment{2$^{nd}$ estimation \eqref{eq_o}}
				 \For{$j = 1,\hdots, b$}	
				 	\State $ X |\widehat{\widehat{S}}_{E}^{(i)}$ \Comment{A time series with PSD $\widehat{\widehat{S}}_{E}^{(i)}$}
					\State $ P(\nu_k;  X |\widehat{\widehat{S}}_{E}^{(i)})$ \Comment{Numerator}
					\State $ X_\ell |\widehat{\widehat{S}}_{E}^{(i)}$ \Comment{$L$ time series with PSD $\widehat{S}_{E}$}
					\State $ \overline{P}_L(\nu_k;  X_\ell |\widehat{\widehat{S}}_{E}^{(i)})$ \Comment{Denominator}
					\State $ \widetilde{P}|\overline{P}_L$ \Comment{Standardized periodogram Eq.\eqref{eq_ptilde0}}
					\State $m^{(i)}_j = \displaystyle{\max_k} ~ \widetilde{P}(\nu_k)\!\!$ \Comment{Max test Eq.\eqref{eq_TM}}	 
				 \EndFor
				 \State $\widehat{P}_{FA}^{(i)}(\gamma; {{\bf m}^{(i)}})$ \Comment{$P_{FA}$ estimate Eq.\eqref{hatpfa}}
 			 \EndFor 
			 \State \textbf{return} $\widehat{\cal{D}}_{\widehat{P}_{FA}}( \{\widehat{P}_{FA}^{(i)}\})$ \Comment{$\widehat{P}_{FA}$ distribution}
 		 \EndProcedure
 	 \end{algorithmic}
\end{algorithm}

\section{Use of GEV distributions}
\label{sec3}
The $\textsf{B}_0$ procedure above does not rely on any model for the cdf of $M$, $ \widehat{\Phi}_{M}$, in \eqref{hatpfa}. To be efficient, this estimation requires $b$ to be large, which makes it computationally expensive (see Sec.\ref{sec5_3}). However, interesting results from univariate extreme-value theory show that the maximum of a set of identically distributed ({dependent} or not) r.v. follows a GEV distribution \cite{Coles_2001}. This suggests that GEV distributions can be used as a model for the cdf of $\max_k\widetilde{P}(\nu_k)$.
This method was used in \cite{Suveges_2012} in the case of white noise but we show below that it can also be used in the case of colored noise when a training data set is available, using the considered standardization.

The GEV cdf depends on three parameters, the location $\mu \in \mathbb{R}$, the scale $\sigma \in \mathbb{R}^{+}$ and the shape $\xi \in \mathbb{R}$ :
	\vspace{-0.1cm}
\begin{equation}
	\label{eq_GEV}
	G(m) := \textrm{Pr}\; (\max_i\{M_i\} \! \le \!m\!) = \mathrm{e}^{\!-\!\Big[ 1+\xi \Big( \frac{{m}-\mu}{\sigma}\Big)\Big]_+^{-\frac{1}{\xi}}},\!
		\vspace{-0.3cm}
\end{equation}
\noindent with $[x]_+ \!:=\! \max(0,x)$. In the case $ \xi \neq 0$, $G$ is only defined for $m : 1+\xi \frac{m-\mu}{\sigma} >0$. In the case $\xi=0$, $G$ is derived by taking the limit of \eqref{eq_GEV} at $0$.
In the previous Algorithm, the unknown parameters ($\xi, \mu, \sigma$) can be estimated using ${\bf{m}}^{(i)}$ (line 12), for instance by maximum likelihood. In this case the maximization can be obtained by an iterative method (see \cite{Coles_2001}).
Denoting by $\widehat{\xi}^{(i)}$, $\widehat{\mu}^{(i)}$, and $\widehat{\sigma}^{(i)}$ the estimated GEV parameters, $\widehat{P}_{FA}^{(i)}$ can be estimated as in \eqref{hatpfa}. The estimated threshold associated to a target $P_{FA}$ can be computed from \eqref{hatpfa} and \eqref{eq_GEV} as:
\vspace{-0.4cm}
$$
	 \widehat{\gamma}^{(i)}(P_{FA})=
	 \hat{\mu}^{(i)} - \frac{\hat{\sigma}^{(i)}}{\hat{\xi}^{(i)}} \Big( 1-\{-{\rm \log}{(1-P_{FA})}\}^{-\hat{\xi}^{(i)}}\Big)
\vspace{-0.1cm}
$$
for $\xi \neq 0$. In the following, we will note by $\textsf{B}^\star$ the $\textsf{B}_0$ procedure using the GEV model.
\vspace{-0.1cm}
\section{Numerical Results}
\label{sec4}
\vspace{-0.1cm}
The first results are obtained using a synthetic colored noise in order to analyze and validate the $\textsf{B}_0$ procedure (Sec. \ref{sec5_2}) and its ``GEV-based'' form (Sec. \ref{sec5_3}). Sec. \ref{sec5_4}
presents an application involving real solar data standardized by hydrodynamic simulations (HDS) of the Sun.
\vspace{-0.1cm}
\subsection{Proposed bootstrap method $\textsf{B}_0$}
\label{sec5_2}
\begin{figure}[h]
\begin{minipage}[b]{\linewidth} \centering
	\begin{picture}(100,100)
		\put(-80,0){\includegraphics[width=\linewidth, height=5cm]{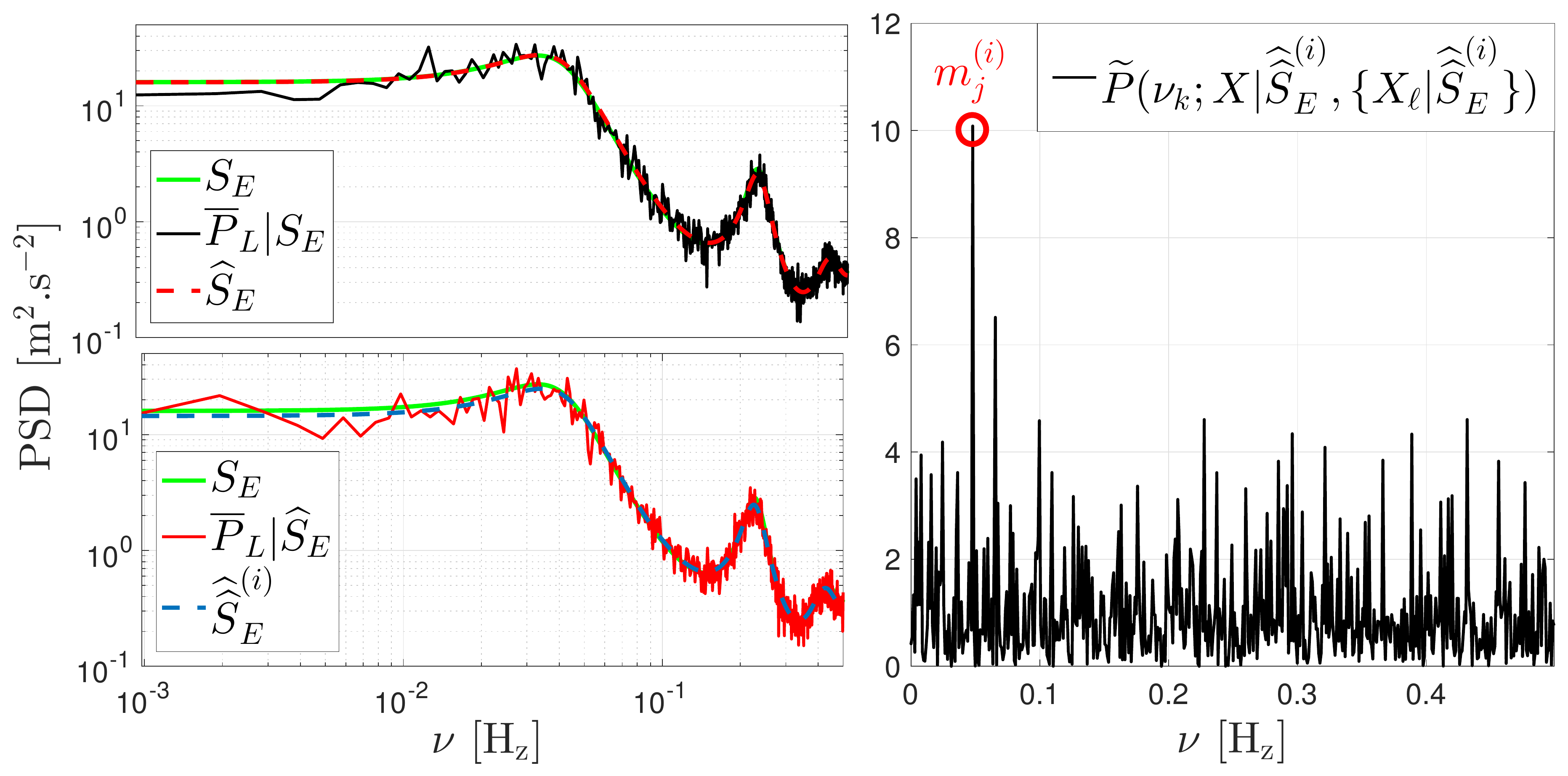} }
		\put(40,128){\bf a) }
		\put(40,68){\bf b) }
		\put(150,105){\bf c) }
	\end{picture}
\end{minipage}
	\hfill
\begin{minipage}[b]{\linewidth} \centering
	\begin{picture}(100,100)
		\put(-80,-20){\includegraphics[width=\linewidth]{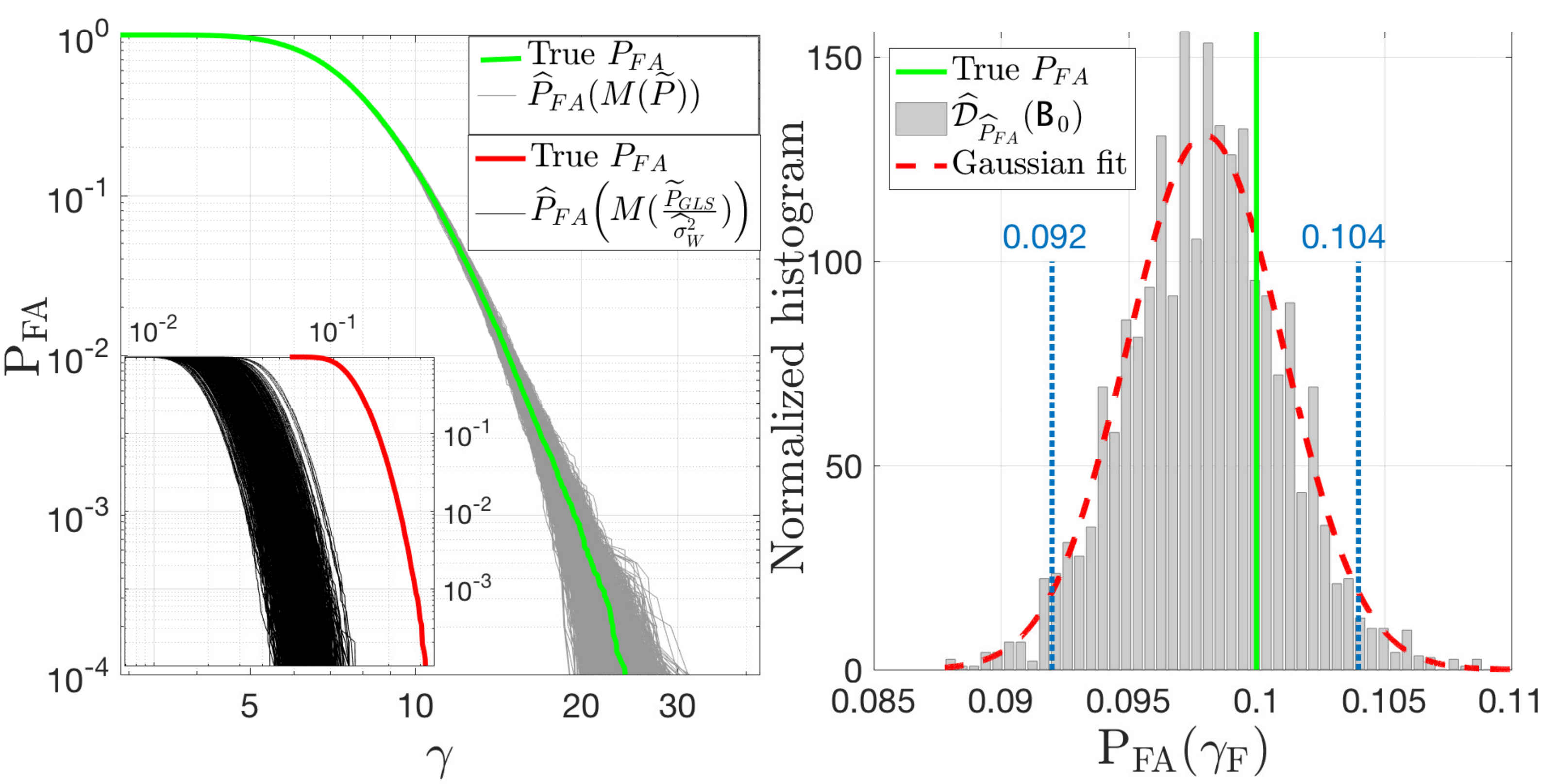} }
		\put(-59,88){\bf d) }
		\put(-59,40){\bf e) }
		\put(150,88){\bf f) }
	\end{picture}
\end{minipage}
\vspace{-1mm}
\caption{Top : Snapshots of the $\textsf{B}_0$ procedure for an AR(6). Bottom : $\widehat{P}_{FA}$ \textit{vs} $P_{FA}$ for several bootstrap methods. }
\vspace{-5mm}
\label{fig1}
\end{figure}
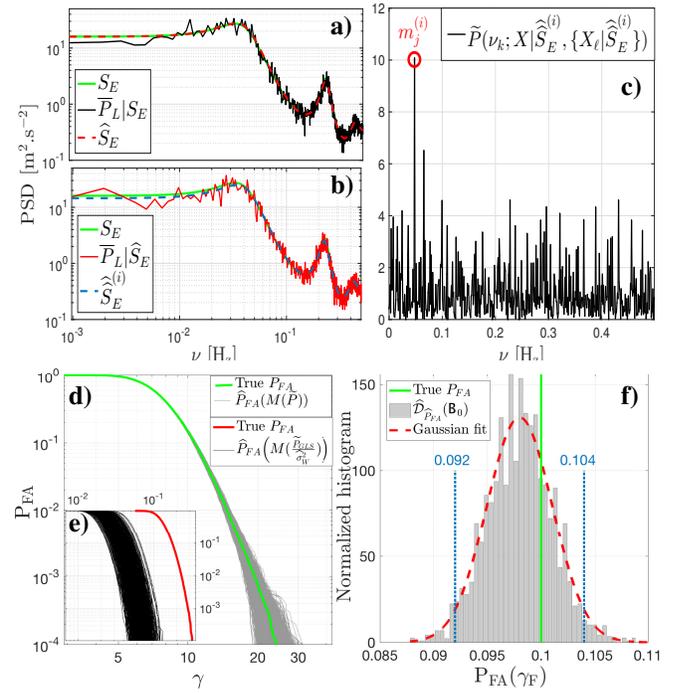

\vspace{-0.1cm}
We first consider a synthetic colored noise, $E$, as an AR(6) process with coefficients $c = \{0.7, 0.05, 0, 0.3, 0, -0.3\}$. The theoretical PSD $S_E$ is given by \eqref{eq_SAR}. It is represented by the green curve in the panels a) and b) of Fig.\ref{fig1}.
For the uneven sampling, we consider a regular grid $\{t_k:=k\,\Delta t\}_{k=1,..,N}$ with $N=1024$ and $\Delta t=1$, from which we randomly select $N=103$ data points.
The first row of Fig.\ref{fig1} illustrates a snapshot of $\textsf{B}_0$ (see Algorithm).
We generate $L=20$ synthetic time series following the true PSD $S_E$, noted $\{X_\ell |S_E\}$, from which we can compute 
an averaged periodogram $\overline{P}_L|S_E$ (panel a), black solid line).
The dashed red curve in panel a) illustrates the primary parametric PSD estimate $\widehat{S}_E$ obtained from $\{X_\ell |S_E\}$ using the $BC$ criterion given in \eqref{eq_o}. 
In panel b), the blue dashed line illustrates one secondary PSD estimate $\widehat{\widehat{S}}_E^{(i)}$ obtained from a `fake training data set' $\{X_\ell |\widehat{S}_E\}$ of $L$ simulated time series. This panel also shows for comparison with a) the averaged periodogram $\overline{P}_L|\widehat{S}_E$ (red solid line).
 Panel c) displays one standardized periodogram obtained from $L+1$ series $ \{X_\ell |\widehat{\widehat{S}}_E^{(i)}\}$, with the maximum value $m_j^{(i)}$ indicated by the red circle.
In this simulation, we obtain $B=5200$ estimates $\widehat{P}_{FA}^{(i)}(\gamma)$ as in \eqref{hatpfa} (grey curves in d). The true $P_{FA}$ (more precisely an accurate estimation obtained through $10^5$ MC simulations of the true AR process) is shown in green. We see that the estimates bound the true $P_{FA}$ over the whole $\gamma$ range.\\
In the inset panel e), we compare with a classical $P_{FA}$ bootstrap estimation procedure that would ignore noise correlations and use the Generalized Lomb Scargle periodogram ($P_{GLS}$) \cite{Zechmeister_2009}, a periodogram frequently used in Astronomy. In this procedure, we generate unevenly sampled
 light curves by resampling the data (through random permutations), evaluate the $P_{GLS}$, compute the maximum and perform test \eqref{eq_TM} with $\widetilde{P}|\overline{P}_L$ replaced by 
 $\widetilde{P}_{GLS}|\widehat{\sigma}_W^2:={P}_{GLS}/\widehat{\sigma}_W^2$ with $\widehat{\sigma}_W^2$ the variance estimated from the data. This provides one estimate $\widehat{P}_{FA}(\gamma)$ and
 we repeat this experiment $10^3$ times. The resulting curves $\widehat{P}_{FA}(\gamma)$ are plotted in black. The
 true $P_{FA}$ of this procedure (in red) has been generated with $10^5$ MC simulations. The method fails to estimate accurately the FA rate, because this simple resampling by permutation breaks the data correlation. This is in clear contrast with the procedure shown in d), which takes benefit from the training data set.
 
\noindent Panel f) shows, in grey, the distribution $\widehat{\cal{D}}_{\widehat{P}_{FA}}(\textsf{B}_0)$ for a threshold fixed at $\gamma_F = 10.6$ corresponding to a true $P_{FA}$ of $0.1$ (green). We see that the estimates bound the true value with a relatively small dispersion. The $95\%$ confidence interval, obtained from the Gaussian approximation (red), is indicated in blue and indeed contains the true $P_{FA}$.

\vspace{-4mm}
\subsection{Use of the GEV distribution}
\begin{figure*}[ht!]
 	 \centerline{\includegraphics[width=0.54\linewidth, height=4.6cm]{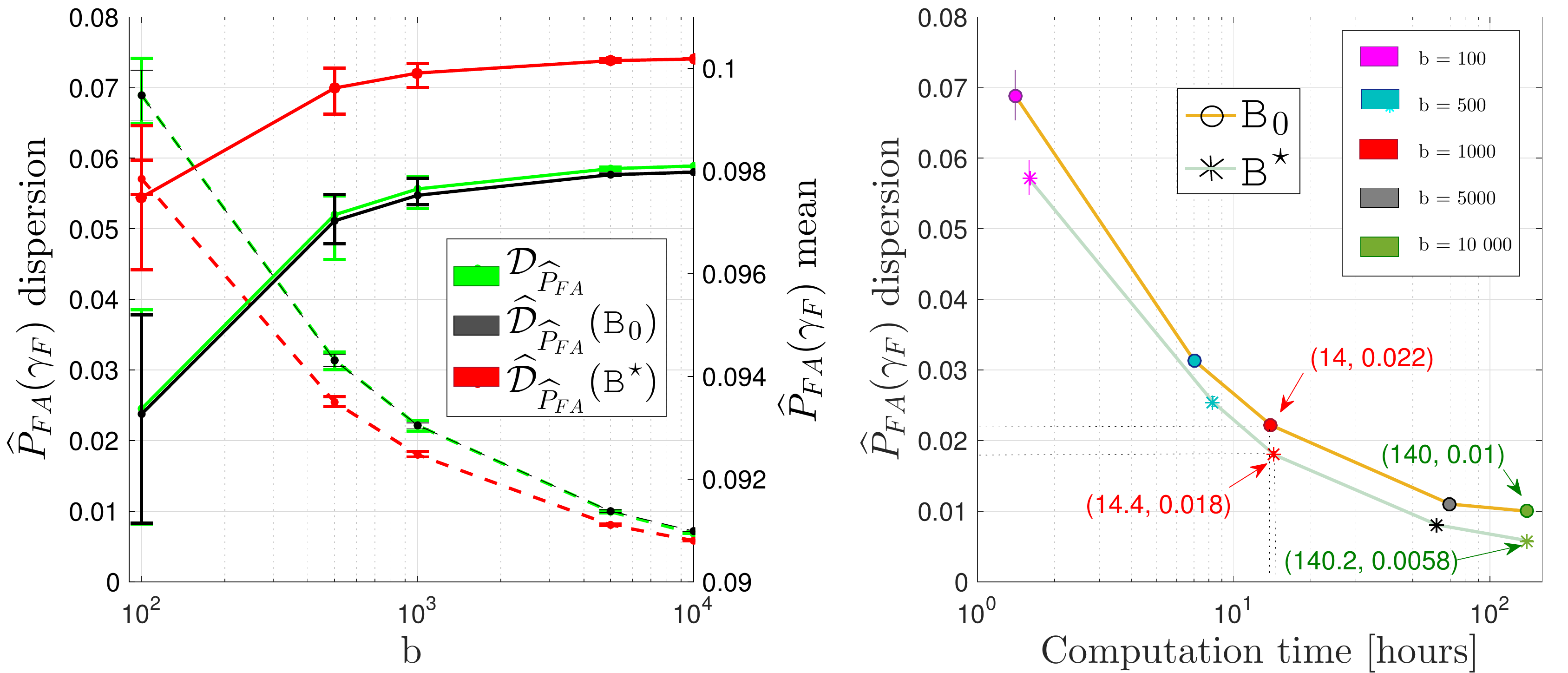}	
	 		   \includegraphics[width=0.54\linewidth,height=4.5cm]{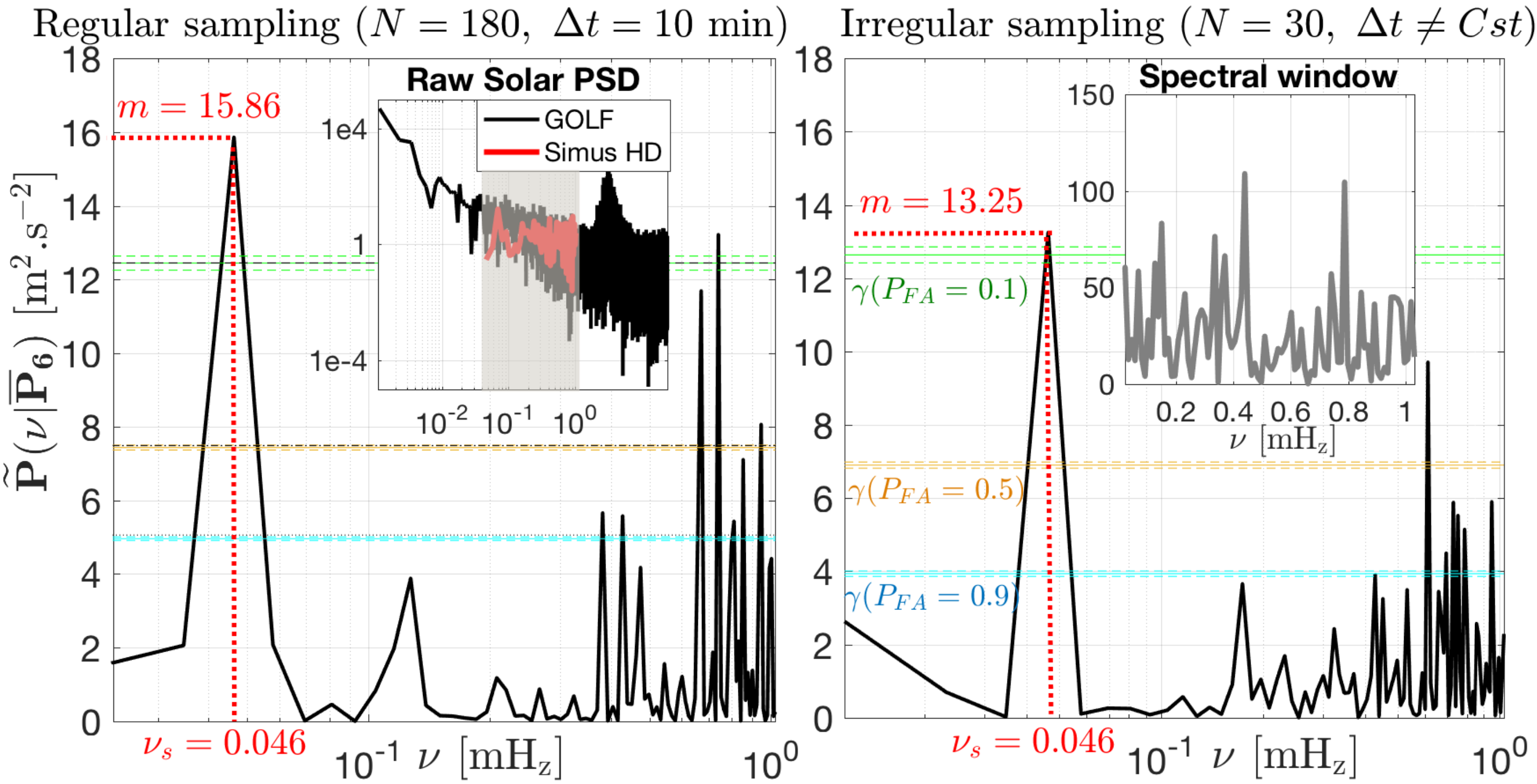}}
  \vspace{-0.3cm}
\caption{Comparison between the $\textsf{B}_0$ and $\textsf{B}^\star$ procedures. ~~~~~~~~~ {\bf Fig. 3.} Illustration of the $\textsf{B}_0$ procedure on real solar data.}
\label{fig2}
\vspace{-0.5cm}
\end{figure*}

\vspace{-2mm}
\label{sec5_3}
In practice, one limitation of the $\textsf{B}_0$ procedure is its computation time, related in particular to the large number ($b$) of periodogram maxima required for the estimation to be accurate. 
To reduce $b$ while keeping a tight confidence interval for the $P_{FA}$, we consider applying the version of the bootstrap procedure using the GEV approximation ($\textsf{B}^\star$).\\
The left panel of Fig.\ref{fig2} compares, as a function of $b$, the empirical means (solid lines, right ordinate axis) and dispersions (dashed lines, left ordinate axis) for the distributions of the FA estimates by three methods (see legend) when the test is ran at a true ${P}_{FA}$ of $0.1$.
 All procedures lead to a decreasing dispersion and bias in the mean
as $b$ increases, with similar estimation performances for ${\cal{D}}_{\widehat{P}_{FA}}$ and ${\widehat{\cal{D}}}_{\widehat{P}_{FA}}$ (respectively in black and green), although for the latter only one genuine set ${\cal{T}}_L$ is available.
Even more interestingly, these results show that the lowest dispersion and bias for the mean $P_{FA}=0.1$ is obtained by the third method ($\textsf{B}^\star$),
because the GEV model is indeed appropriate and has far less degrees of freedom than the non parametric estimate $\widehat{\Phi}_M$ used in the two other bootstraps.
%
The right panel compares the possible compromises dispersion \textit{vs} computation time for $\textsf{B}_0$ and $\textsf{B}^\star$. We see that $\textsf{B}^\star$ allows for better compromises, with a $\approx 20-40\%$ lower dispersion than $\textsf{B}_0$ at the same computational cost.
\vspace{-3mm}
\subsection{Application to exoplanet detection}
\vspace{-1mm}
\label{sec5_4}
\noindent As in \cite{Sulis_2017a}, an application of this work is the detection of exoplanetary signatures in the noise of RV data. 
In short, orbiting planets create, through the gravitational force, a quasi-periodic Doppler shift in the stellar light, leading to so-called RV time series. 
The stellar surface of solar-like stars are convectively unstable. Convection is a stationary stochastic process which generates (large) fluctuations around the hydrostatic equilibrium of the star.
HDS of the inhomogeneous pattern (called granulation) have been substantially improved in the last decade \cite{Nordlund_2009} offering the opportunity to generate simulated time series (yet few, owing to the computational cost of the HDS), that can be used as a training data set ${\cal{T}}_L$. 
	\vspace{-0.2cm}
\vspace{0.1cm}

\noindent We consider here solar RV data acquired by the GOLF spectrophotometer on board SoHo satellite \cite{Garcia_2005}.
The periodogram of 30h observations is shown in the inset panel of Fig.3 (top, black). 
Different phenomena occur depending on the considered frequency. Here, we focus only on the {granulation noise}, which is active over minutes to days (grey shaded region) corresponding to the temporal scale of close-in planets. As an illustration of HDS, we show an averaged periodogram $\overline{\bf{P}}_L$ obtained through $L=6$ independent HDS of the granulation solar noise \cite{Bigot_2008}, covering the frequency range associated to periods of $10$ min to 1 day (red).

\noindent For our experiment, we introduce in the data a fake periodic signal: $s(t) \!=\!\! A \sin(2\pi t \nu_s)$, with $A\!=\!0.28$ m.s$^{-1}$, $\nu_s \!=\!0.046$ mHz (for these data, $ \frac{A}{\sigma_{X_{obs}}} \approx0.22$). 
This situation corresponds to a Keplerian signature of a $0.27$ Earth-mass planet, orbiting a Sun-like star with circular orbit and period of $6$h.\\
 The left panel of Fig.3 displays $\widetilde{P}$ obtained from evenly sampled data (see title). In this case the $P_{FA}$ of test \eqref{eq_TM} is given Eq.(18) of \cite{Sulis_2017a}. The theoretical (asymptotically exact) thresholds corresponding to three target $P_{FA}$ (respectively, $0.1$, $0.5$ and $0.9$) are indicated by the black horizontal (respectively dotted, dash-dotted and dashed) lines. The color dashed lines indicate the $\approx 95\%$ confidence intervals obtained by $\textsf{B}_{0}$ for the three thresholds. Comparing with the black lines shows that the bootstrap method works well.
 Similar results are obtained in the case of uneven sampling (right panel), where the data is obtained by randomly selecting $N=30$ samples of the evenly sampled time series. 
 The corresponding spectral window (modulus of the Fourier Transform of the observation window) is plotted in grey in the inset panel. 
\vspace*{-0.3cm}
\section{Conclusions}
\vspace*{-0.1cm}
This paper has investigated the possibility of using noise training data sets to improve the control of the false alarm rate associated to detection tests. 
This method is based on standardized periodograms, which have been considered in a previous study \cite{Sulis_2017a} but in the case of regularly sampled observations. 
By developing an adapted bootstrap procedure in the Fourier domain, it appears that one can reliably bound the false alarm probability in the general case of irregularly sampled observations. 
However, as this procedure is based on MC simulations, it is computationally heavy. Exploiting a GEV distribution allows to save time while keeping the same behavior for the resulting FA estimates. This procedure can be useful for the detection of extrasolar planets in RV data. In this case, the time sampling is often uneven, the exoplanetary signatures hidden in the colored noise coming for the stellar surface convection and this noise can be simulated using astrophysical codes.\\

\vspace*{-0.3cm}
{\bf \noindent Acknowledgement} 
\textit{ We are grateful to Thales Alenia Space, PACA r\'egion and Programme National de Physique Stellaire of CNRS/INSU, for supporting this work. And, we thank
the the GOLF team for providing  the solar time series.
}



\footnotesize
\bibliographystyle{IEEEbib}
\bibliography{Biblio_Eusipco} 



\end{document}